# A High-Temperature Thermocouple Development by Additive Manufacturing: Tungsten-Nickel (W-Ni) and Molybdenum (Mo) Integration with Ceramic Structures

[1]Azizul Islam, [1]Aayush Alok, [1]Vamsi Borra, [1]Pedro Cortes

[1]Advanced Manufacturing Research Center, Youngstown State University, OH 44555

## *Abstract*

Additive manufacturing holds more potential to enable the development of ceramic-based components. Ceramics offer high resistance to heat, high fracture toughness, and are extremely corrosion resistant. Thus, ceramics are widely used in sectors such as the aerospace industry, automotive, microelectronics, and biomedicine. Using various additive manufacturing platforms, ceramics with complex and uniquely designed geometry can be developed to suit specific applications. This project aims at innovating high-temperature thermocouples by embedding conductive metal pastes into a ceramic structure. The paste used includes tungsten, molybdenum, and antimony. The metal pastes are precisely extruded into a T-shaped trench inside the ceramic matrix. Following specific temperature ranges, the ceramic matrix is sintered to improve the properties of the material. The sensors produced can function at extremely high temperatures and are thereby suitable for high-temperature environments. Comparative testing of the 3D sintered sensors with conventional temperature sensors shows high correlation between the two classes of sensors. The resulting R-squared value of 0.9885 is satisfactory which implies the reliability and accuracy of 3D sintering sensors are satisfactory in temperature sensing applications.

## Introduction

Ceramics has attracted considerable interest in the field of materials science due to its wide range of applications such as dental implants, electronics, chemical, refractory, aerospace, and automotive in a variety of industries [1][2][3][4][5]. Features such as heat resistivity, electrical resistivity, corrosion resistance, and high fracture strength make them exemplary candidates for the industries. Simultaneously, electronics that can be employed in high temperatures are also garnering interest. Parts with ceramic outer shell can be beneficial in such instances. However, manufacturing intricate ceramic parts is difficult in terms of expenditure, and effort. Additive manufacturing (AM) allows the production of complex and intricate structures to address unique and customized designs for the ease in layer-by-layer fabrication [6][7]. 3D printed ceramics have been produced using all the seven AM technologies established by the ASTM for internally designed geometries [8][9].

AM techniques were previously used to produce ceramic core and shell molds for various applications in architectures, consumer products in electronics, dental and automobile, and aerospace industries[10]. Smart materials embedded in ceramic substrates would result in various advanced applications for their notable features. Sensors that can operate in extreme environments have abundant applications in aircraft, turbine engines, aerospace systems, and material processing systems. Due to their high heat resistivity and dielectric properties, ceramic materials have been studied for the use of high-temperature sensors[11]. Zhao et al[12] implemented a poly-derived ceramic (PDC) material composed of silicon aluminum carbonitride to produce a high-temperature sensor capable of recording temperatures up to 830°C. The sensor used embedded Pt wires to act as electrodes, while the PDC sensor collected accurate temperature measurements for a span of

10 hours. The resulting high-temperature sensor proved to have excellent accuracy and repeatability when compared to a commercial thermocouple.

Hossain et al [13] researched manufacturing a smart sensor by a multi-step process using EBM of a metal shell and a ceramic piezoelectric sensor core. Corson et al [14] detailed the usage of various AM techniques for ceramic materials that can be utilized in energy applications. The study shows the importance and growth of using engineering ceramics for thermoelectric, thermoionics and carbon capture and storage. This review also highlights the limitation in the design and manufacturing techniques. Choi et al [15] exhibited a study of encapsulate a proximity sensor in a ceramic housing. The encapsulated sensor was able to precisely capture the inductances at elevated temperatures. Hoerber et al [16] studied powder -based additive manufacturing technique to print PMMA shells and aerosol-jet printing for establishing electronic connections on the sample for electronic applications. Although the connections show that adaptions for assembly techniques, this technique cannot be used for integrated circuits or mass production. The lack of one straight method for production for smart ceramic materials is always seen which can be a serious economic loss for industries such as aerospace and refractories.

Given the difficulty in embedding multi-materials with deviations involved in ceramic manufacturing and production techniques, a cross-layer integration by AM can benefit in production of these smart materials. This can be achieved by simultaneous AM printing of sensors and ceramic housing or printing them separately using AM techniques (hybrid model) [17]. The present study focused on developing a better way to produce embedded sensors majorly for high temperature applications. A hybrid model using two different AM techniques was established to produce the sensor as a proof-of-concept. A ceramic substrate and a conductive ink were separately printed. The fabrication technique used in this study involves a multi-step process to produce embedded sensors using AM techniques. A first step includes producing ceramic substrate using SLA, DLP, and BJP, and FDM technique to deposit conductive metal oxides on the printed ceramic substrates.

VPP is a well-known, cost-effective additive manufacturing technique used to produce ceramic components [18][19][20]. VPP uses an ultraviolet light beam to selectively cure a photopolymer slurry with ceramic suspension in a layer-by-layer fashion [21]. A VPP printer operates by use of UV light where the build platform lowers into the resin tank and the UV light from beneath reacts with the resin, resulting in the cross-linkage of monomers. These cross-linked monomers bind together to form polymers. This process occurs in microlayers, and results in the formation of a solid, 3D object. VPP provides good printing accuracy and resolution, as well as better mechanical properties and surface finishes as compared to other printing techniques [22][23][24]. StereoLithography (SLA) and Digital Light Processing (DLP) are two different classes of VPP which vary in terms of resolution and accuracy of printing [25]. BJP is a powder-based 3D printing process that creates a 3D part by employing a liquid binder on thin layers of powder, adhering the powder layers into a uniform solid part[26][27][28]. Advantages of this technique include vast material variety printing and large part production. Material Extrusion is one of the AM processes which continuously extrudes material through a nozzle and subsequently solidified finally to produce a 3D part.

The new model offers a new method to produce thermal sensors, using metal inks as the sensing element and 3D-printed ceramic as a substrate with a T-shaped trench that protects and encloses metal inks. Additively produced ceramics with a high-temperature capability and two kinds of metal inks, Tungsten-Nickel and Molybdenum, use Seebeck effect to detect the temperature accurately. The sensors and the thermocouples under thermal stress have proven successful in maintaining the intricate geometries of the thermocouples and the ceramics even after the experiment. It indicates that this breakthrough is highly useful in the engineer's application in harsh high-temperature environments.

# Methodology of the Study

This study, therefore, investigates the structure's properties and functionality of the ceramic sensor by using a range methodology that entails multiple key steps. Among the initial steps entails the design and fabrication process of the ceramic structure. Additionally, the next step entails the preparation and application of specific inks that are used and lastly the sintering process. After the preparatory stages capture of relevant sensor outputs was carried out through data acquisition. In the last case, the sensor's performance was tested against the established benchmarks through the comparison test after the wire connection. This methodology was used to make a comparison test of the established benchmarks and the sensor.

I. Ceramic Structure Design and Fabrication:

The first step of this methodology is to develop a ceramic structure of extreme reliability. In this study, this was developed in Autodesk Fusion 360. The extreme dimensions of 25mm x 25mm x 4mm, this cuboid has a 2mm diameter trench to create a T shape when viewed from the top. This is due to the hole in the middle of the structure, which will ensure the inks overlap each other. Figure-1 shows the configuration, and it would be achieved with high precision to ensure maximum functional properties of the sensor.

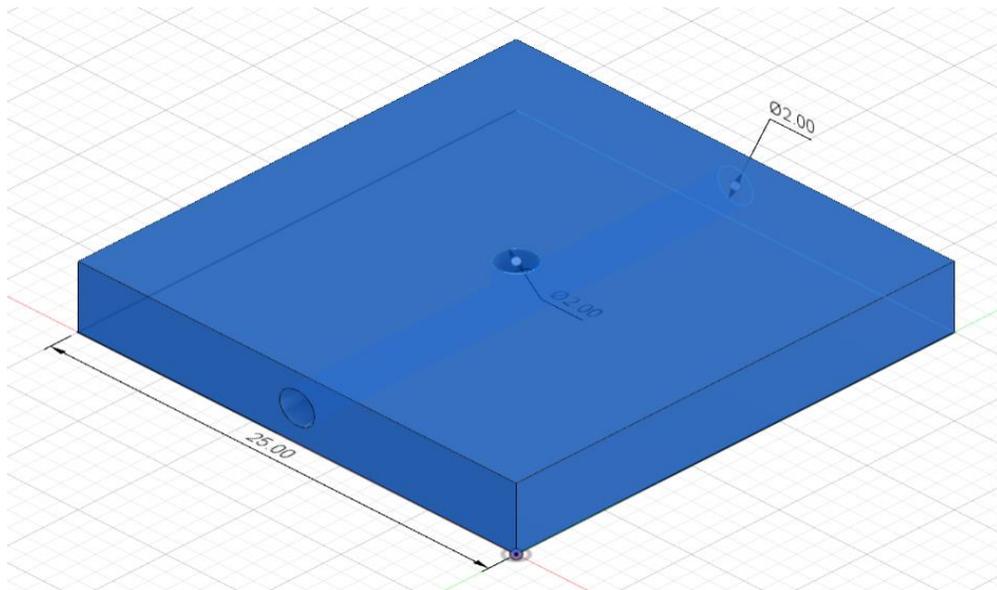

**Figure 1.** CAD Model of Ceramic Structure of Sensor

After design, the foundation of this research involved fabricating 3D-printed ceramic high-temperature sensor samples with good dielectric properties. The ceramic substrate materials include alumino-silicates, fused silica, alumina, and zirconia. These materials were selected for their advanced properties, including their stability at high-temperatures, mechanical robustness, and stable electrical properties. The favorable characteristics of alumina include both stable electrical and mechanical properties. Fused silica is one of the inexpensive abundantly available heat resistant and laser resistance material that can be used for aerospace applications. Alumina is resistant to wear from thermal and corrosive stress and can withstand high temperatures and thermal shock [4]. Zirconia can resist corrosion and has high surface area and thermal shock resistance. The electrical and thermal properties of zirconia make the ceramic ideal for circuits and optics [29].

To fabricate the ceramic part, three additive manufacturing processes were examined in this study. These techniques include Binder Jetting Printing (BJP), StereoLithography (SLA), and Digital Light Processing (DLP).

a) **Binder Jetting Printing:** ExOne Innovent 3D printer was used to print alumino-silicates ($d_{90}$ = 70 µm) using an aqueous binder (BA#0040). A layer thickness of 75µm was used to print the samples with trenches. The hopper containing powder uses ultra-sonic vibrations to drop the powder on the build platform and a roller moves on the powder layer to level the surface. The print head selectively drops 40 pL binder drops selectively on each single layer. The samples printed were cured at 600°C for 6 hours before the samples were removed from the supporting system. Later the samples are stacked with their counterparts to create an overmold and sintered together. The samples are sintered at 1275°C (Figure 2(a)).

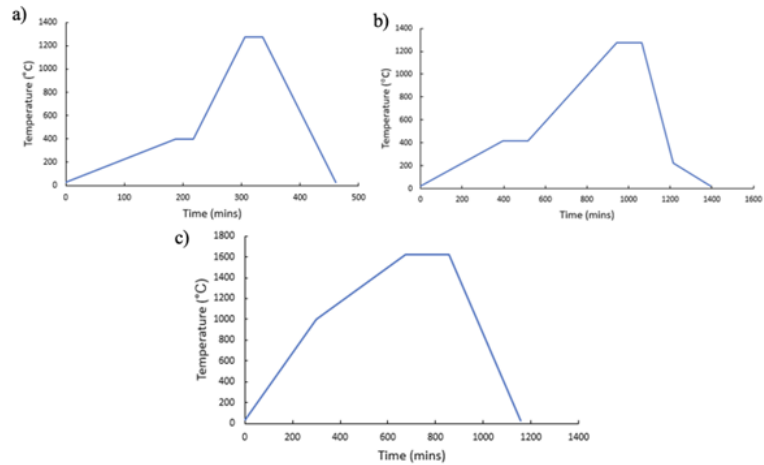

**Figure 2.** The sintering profile for FormLabs silica involves a ramp to 420°C followed by a two-hour dwell, then another ramp up to 1275°C. The ceramic part dwells for another two hours and is then cooled.

b) **StereoLithography**: Commercially available silica resin (Formlabs, Somerville, MA) was used to print the samples on Formlabs Form 2 unit. The unit contains a 405nm UV light to cure the ceramic layers onto the build platform. The samples were printed in a bottom-up approach with sliced layers and support structures. After the samples are completely built they are separated from the build platform and washed in FormWash (Isopropyl alcohol) to remove excess resin. These green parts are then sintered in a box furnace at a temperature reaching 1275°C (Figure 2(b))

c) **Digital Light Processing:** Admatec Admaflex 130 DLP printer was used to obtain high resolution alumina parts. This printer contains a reservoir for slurry-holding and a rolling foil base in which the slurry is carried from the reservoir to the build platform. Once the rolling foil carries the slurry under the build platform, the platform lowers, and an LED light cures the slurry selectively onto the build platform to produce high-resolution parts. The foil then carries the excess slurry to a pump, where the slurry is pumped back to the reservoir and recycled for efficient feedstock management. Once the parts are printed, they endure a 24-hour water debinding process in DI water. They are then thermally debinded and sintered up to 1650°C to produce fully dense ceramic parts (Figure 2(c))

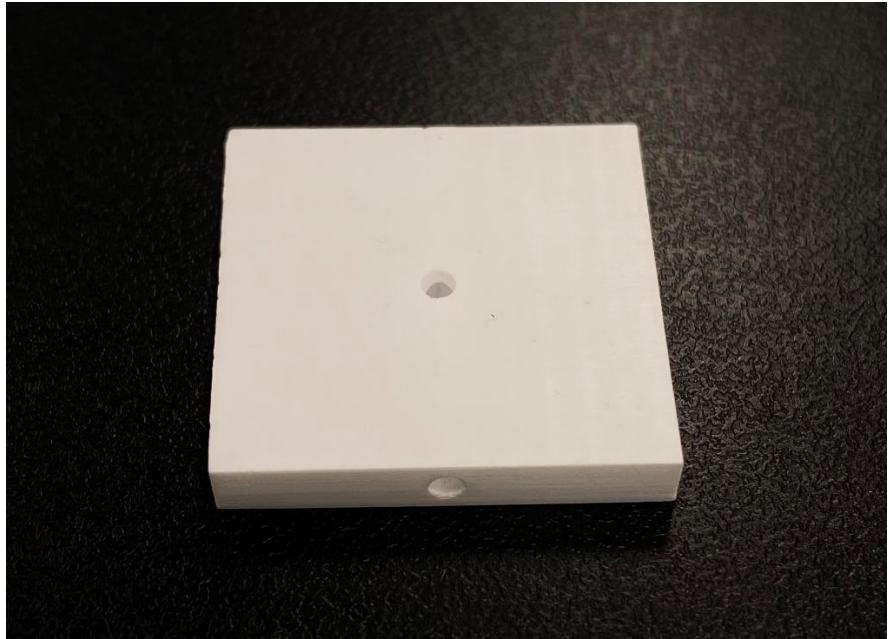
**Figure 3.** Sintered SLA silica part

II.     Preparation and extrusion of Inks:

Following the fabrication of the final Sintered SLA silica part, a tungsten composite filament was created in the laboratory. The tungsten composite used in the research was provided by MIT Lincoln Labs and contained tungsten particles ranging from 1-1.5 micrometers in diameter. The paste was created in the lab using 52 wt. % of SIS polymer and 48 wt. % of pure toluene. These components were added to a glass vial and then mixed with a vortex mixer. Once a gel-like consistency was formed, the tungsten powder was added, and the vial was mixed again with the electric mixer. The tungsten filament showed significant electrical properties, including conductivity and resistance with CTE of 4.6 ppm/°C. However, the CTE is increased by addition of Ni to the tungsten [30]. Along with, a second conductive paste was also implemented in this research, molybdenum paste was provided by Applied NanoTech Inc. The paste showed semi-conductive properties in the green state and gained conductivity after being sintered in an inert gas to temperatures exceeding 1100°C. The thermal coefficient of expansion for the Mo paste is approximately 4.8ppm/°C.

The HyRel 3D printer was employed to prepare and deposit tungsten-nickel and molybdenum inks meticulously before and after the process. A tylo surpassed trough was prepared in which the Moly ink was first extruded. For optimal viscosity and flow properties, the ink was first warmed to 65 degrees Celsius over the span of ten minutes. Similarly, the W-Ni ink was extruded so that the interface between two distinct inks was clearly visible. Careful note had to be taken of the point where the two inks contacted, which was in the center of the tylo verged trough presented in the figure to establish a continuous and uniform junction that was critical for structural integrity. With the inks properly deposited and aligned, the assembly was ready to move on to the next steps.

III.    Sample Sintering:

Having prepared the sample containing a deposition layer of tungsten-nickel and molybdenum, which are the key metals to be sintered, we entered the most crucial stage that secured the essential mechanical and thermal properties of the structure. In the initial step, a kinesimeter purged the furnace with inert argon gas at an 80 ml/min flow rate to eliminate all oxygen from the heating area to prevent metal pastes from oxidation. The temperature began to increase gradually from 25°C to 200°C at 0.2 °C a minute regime to prevent rapid thermal expansion that could lead to different material deformations. Afterward, we increased the temperature of the process to 1100°C at a 0.41 °C a minute regime to ensure that the furnace heated evenly and there was a uniform diffusion of W-Ni and Mo particles for higher sintered body strength. As the temperature reached 200°C, we held this constant for 120 minutes to ensure that all binders burned out, and the metals were properly heated for welding. We kept the sample in a continuous 1100°C regime with the abovementioned increasing regime for another 120 minutes to ensure that all particles were fused on a molecular level and began cooling to 25°C at 4 °C a minute regime for the stabilized microstructure through minimal thermal stresses. The entire process lasted 3578.75 minutes, or 59.65 hours, to acquire optimal values representative of high-quality W-Ni-Mo composites. Table-1 indicates the total steps of sintering.

**Table 1:** Temperature profile of sintering after Appling tungsten-nickel and molybdenum

| | |
|---|---|
| Step 1: Ramp | 25°C to 200°C at the rate of 0.2/min |
| Step 2: Dwelling | Hold 200°C temperature for 120min. |
| Step 3: Ramp | 200°C to 1100°C at the rate of 0.41/min or 25°C /hr. |
| Step 4: Dwelling | Hold 1100°C temperature for 120min. |
| Step 5: Cooling down | From 1100°C to 25°C at the rate of 4°C /min. |

IV.     Electrical Connections:

To be able to record and collect real temperature data, the above-prepared high-temperature sensor had to have electrical connections established. Specifically, we used high-temperature K-type thermocouple wire specially designed for temperature data collecting in high temperatures. The wire is comprised of Chromel and Alumel, making it thermoelectric and capable of measuring a wide range of temperatures with a standard accuracy of ±2.2 °C or ±0.75%. It is manufactured as a 24 AWG solid wire, meaning that the bare wire's actual diameter was 0.51 mm with high-temperature fiberglass insulation to prevent from melting up to 700 °C. The insulation has ANSI-approved color shading, which indicated red color to be negative and yellow to be positive. The connecting point for yellow Chromel took place on a W-Ni paste, which serves as the cathode while connecting the red Alumel took place at a molybdenum paste, which served as an anode. To secure the wire and prevent its cracking during exposure to high temperatures, we used specially designed glue with high-temperature absorbing capabilities. Graphically, in the documentation, this fourth step may be illustrated using Figure- 4, displayed the sample after this step was completed, and it is ready for operational testing.

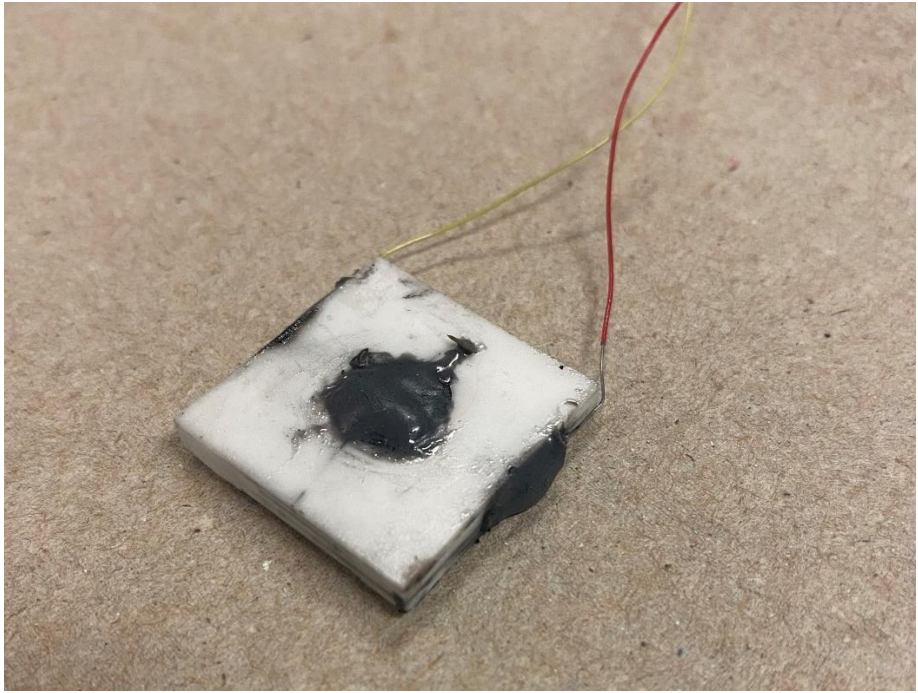

**Figure 4.** Image of sensor after wire connection and applying glue.

V.    Data acquisition and Comparison Test:

During the phase of data acquisition and comparison testing, we combined the use of the conventional temperature sensor and our engineered one, and to perform the comparison test, both sensors were used in conjunction with a DAQ970A Data Acquisition System by Keysight. Critical to temperature monitoring and recording, we set up the samples and sensors within a furnace to enhance the accuracy of measuring temperature. We utilized a conventional Type-K thermocouple for benchmarking purposes as the reference sensor to be used across the experiment to facilitate the data collected by the reading we acquired from the innovative sensor.

Our data collection began by taking readings on the relationship between temperature and voltage, focusing on the differences in voltage output readings from our new sensor as we increased the temperature from the room temperature of 25 degrees Celsius to 165 degrees Celsius. This can be considered the very first Temperature Vs Voltage data collection, as it helped develop the sensor response curve that creates a basis.

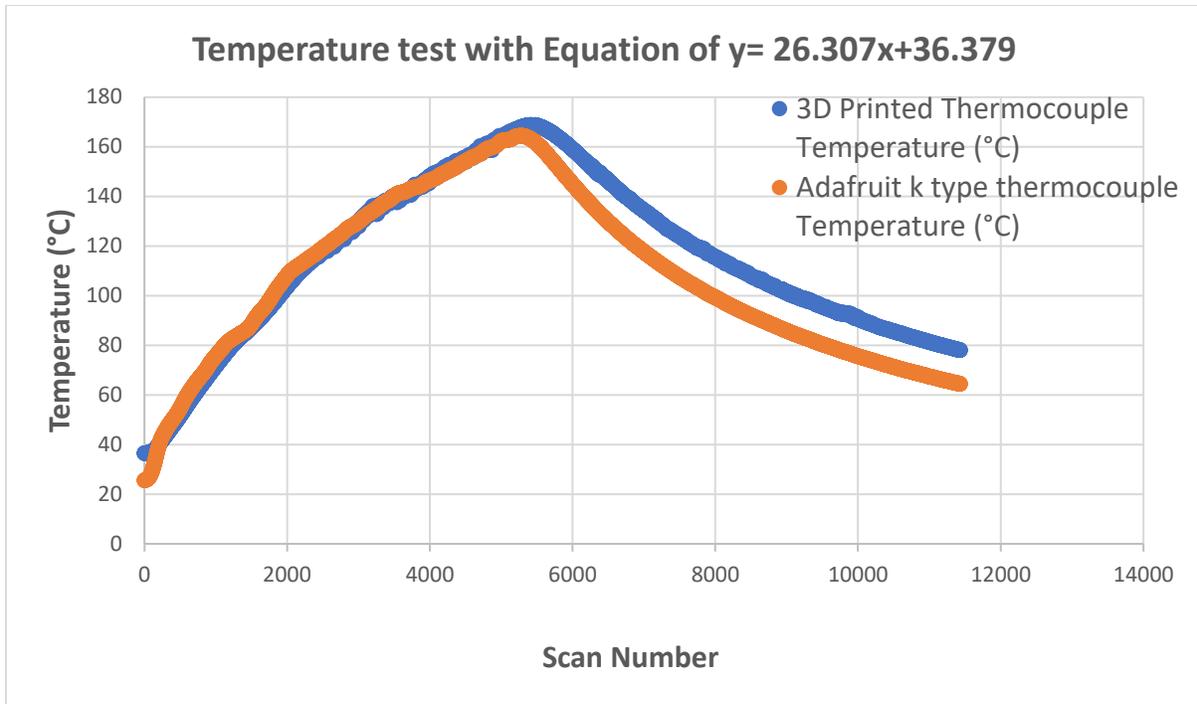

**Figure 5.** The Curve shows the data of Temperature comparison test between 3D printed thermocouple and Adafruit K-type thermocouple.

After the collection of preliminary data was completed, we continued with the comparison test using the same setup described above. Such a test was especially impactful as it allowed us to determine how the performance of our novel sensor correlates with that of the traditional Type-K thermocouple in the same experiment. This way, any variations in the received data could be conclusively attributed to the sensors' designs and materials. This highly comprehensive approach confirmed not only the efficiency of the new sensor employed in our experiment but also the accuracy and reliability of the read data.

## Result and Discussion

The results obtained from this data collection experiment, Temperature Vs Voltage, provided proof of correlation between the temperature increment and the voltage output of the newly formulated sensor. The sensor started by giving the smallest voltage output when it was at room temperature at 25 degrees Celsius, recording 0.003 mV. The output voltage kept on increasing dynamically as the temperature was being increased, and at 165 degrees Celsius, it recorded 5.009 mV. This data was adequate in providing a picture of the sensor's sensitivity to the range of temperature covered.

Based on the collected data, we derived a linear regression equation to quantify the relationship between temperature and voltage for the sensor:

$$\text{Temperature} = 26.307 \times (\text{Voltage in mV}) + 36.379$$

This equation provided a mathematical expression of the sensor's output that could be used to accurately predict the temperature given a certain voltage.

Finally, we validated our sensor efficacy by performing a comparison test with the above equation. This comparison enabled us to quantify the performance of our sensor through the calculation of the coefficient of determination, represented as R-squared value. The computed R-squared value, 0.9939 implied that 99.39% of the variance in the measured temperatures could be attributed to that of voltage output predicted by our model. The user obtained a high R-squared value, which shows that it procures a high linear correlation; hence, its behavior is highly sensitive and stable.

The outcomes of the experimentations reinforce the ability of the sensor to provide accurate and precise temperature readings and support its adoption in systems that demand precise temperature monitoring. The evidence presented by the linear regression analysis and the high value of R-squared exhibits the robust nature of the sensor and opens it up to industrial and research environments for enhanced temperature supervision. In this regard, the results do not only authenticate the design of the sensor but also outmatch other traditional temperature sensors to revolutionize the temperature measurement field.

# Conclusion and Future Work

This study has indicated the possibility and efficiency of utilizing the additive manufacturing method for producing high-temperature thermocouples entrenched into the ceramic body. The use of the conductive elements, namely tungsten-nickel and molybdenum pastes, within the 3D-printed ceramic matrix has facilitated the development of the sensors that can function under high-temperature conditions in a precise and reliable manner. This is corroborated by the R-squared coefficient when contrasting the outcomes of test comparison, which amounted to 0.9939. It evidences the stability of functioning and high level of precision in the sensors as opposed to the traditional thermocouples. In this way, the present research supports the feasibility of additive manufacturing for generating intricate and reliable ceramic formations critical for high-temperature industries such as aerospace, automotive, and others.

In what concerns the prospects for the future, extending the varieties of materials and further optimization of the design may be the key for the improved performance of such sensors. For example, other metal-ceramic composites or ceramics of another type may be utilized to increase the temperature at which the sensors function and introduce broader possibilities for application. Still, it is essential to note that it is necessary to use the developed computational models to predict and optimize thermal and mechanical behavior during operation first. At the same time, implementing methods to scale the production process of such devices for industrial use would require the improvement of additive manufacturing techniques in terms of efficiency and costs.

Future research would also venture into integrating these sensors into more comprehensive systems, i.e., turbine engines or spacecraft, to enable effective high-temperature process monitoring and control. Accordingly, future works would necessitate developing compatible electronics and data acquisition systems to be used at high temperatures. More so, integrating smart features into the sensors, such as real-time data processing and wireless data transmission, would input them as part of IoT systems in industrial applications. Fulfilling these aspects shall trigger the implementation of 3D-printed ceramic-based sensors in other high-temperature environments beyond the scope of the current study. These would revolutionize the high temperature monitoring sector, significantly contributing to material science and additive manufacturing processes.

# References

[1] S. Manotham, S. Channasanon, P. Nanthananon, S. Tanodekaew, and P. Tesavibul, "Photosensitive binder jetting technique for the fabrication of alumina ceramic," J Manuf Process, vol. 62, pp. 313–322, 2021.

[2] G. Miao, W. Du, M. Moghadasi, Z. Pei, and C. Ma, "Ceramic binder jetting additive manufacturing: Effects of granulation on properties of feedstock powder and printed and sintered parts," Addit Manuf, vol. 36, p. 101542, 2020.

[3] B. Mummareddy, E. Burden, J. G. Carrillo, K. Myers, E. MacDonald, and P. Cortes, "Mechanical performance of lightweight ceramic structures via binder jetting of microspheres," SN Appl Sci, vol. 3, pp. 1–10, 2021.

[4] J. A. Gonzalez, J. Mireles, Y. Lin, and R. B. Wicker, "Characterization of ceramic components fabricated using binder jetting additive manufacturing technology," Ceram Int, vol. 42, no. 9, pp. 10559–10564, 2016.

[5] A. Islam, S. E. A. Himu, N. Bin Amin, M. S. Kaoser, F. Abrar, and M. R. Islam, "Pulmonary Artery Pressure (PAP) Prediction Based On Three Physiological And Seven ECG Signal Features," in 2022 IEEE 3rd Global Conference for Advancement in Technology (GCAT), IEEE, 2022, pp. 1–5.

[6] S. Y. Song, M. S. Park, D. Lee, J. W. Lee, and J. S. Yun, "Optimization and characterization of high-viscosity ZrO2 ceramic nanocomposite resins for supportless stereolithography," Mater Des, vol. 180, p. 107960, 2019.

[7] Y. Lakhdar, C. Tuck, J. Binner, A. Terry, and R. Goodridge, "Additive manufacturing of advanced ceramic materials," Prog Mater Sci, vol. 116, p. 100736, 2021.

[8] M. Revilla-León, M. J. Meyer, A. Zandinejad, and M. Özcan, "Additive manufacturing technologies for processing zirconia in dental applications," Int J Comput Dent, vol. 23, no. 1, pp. 27–37, 2020.

[9] H. Hassanin, K. Essa, A. Elshaer, M. Imbaby, H. H. El-Mongy, and T. A. El-Sayed, "Micro-fabrication of ceramics: Additive manufacturing and conventional technologies," Journal of advanced ceramics, vol. 10, pp. 1–27, 2021.

[10] H. Bikas, P. Stavropoulos, and G. Chryssolouris, "Additive manufacturing methods and modelling approaches: a critical review," The International Journal of Advanced Manufacturing Technology, vol. 83, pp. 389–405, 2016.

[11] Z. Ren, S. Bin Mujib, and G. Singh, "High-temperature properties and applications of Si-based polymer-derived ceramics: a review," Materials, vol. 14, no. 3, p. 614, 2021.

[12] R. Zhao, G. Shao, Y. Cao, L. An, and C. Xu, "Temperature sensor made of polymer-derived ceramics for high-temperature applications," Sens Actuators A Phys, vol. 219, pp. 58–64, 2014.

[13] M. S. Hossain et al., "Fabrication of smart parts using powder bed fusion additive manufacturing technology," Addit Manuf, vol. 10, pp. 58–66, 2016.

[14] C. L. Cramer et al., "Additive manufacturing of ceramic materials for energy applications: Road map and opportunities," J Eur Ceram Soc, vol. 42, no. 7, pp. 3049–3088, 2022.

[15] J.-W. Choi, R. Huang, A. Urban, D. Jiao, and J. Zhe, "Inductive Proximity Sensors within a Ceramic Package Manufactured by Material Extrusion of Binder-Coated Zirconia," Available at SSRN 4038100.

[16] J. Hoerber, J. Glasschroeder, M. Pfeffer, J. Schilp, M. Zaeh, and J. Franke, "Approaches for additive manufacturing of 3D electronic applications," Procedia CIRP, vol. 17, pp. 806–811, 2014.

[17] D. Lehmhus et al., "Customized smartness: a survey on links between additive manufacturing and sensor integration," Procedia Technology, vol. 26, pp. 284–301, 2016.

[18] M. L. Griffith and J. W. Halloran, "Freeform fabrication of ceramics via stereolithography," Journal of the American Ceramic Society, vol. 79, no. 10, pp. 2601–2608, 1996.

[19] T. Chartier, C. Chaput, F. Doreau, and M. Loiseau, "Stereolithography of structural complex ceramic parts," J Mater Sci, vol. 37, pp. 3141–3147, 2002.

[20] E. Zanchetta et al., "Stereolithography of SiOC ceramic microcomponents.," Adv Mater, vol. 28, no. 2, pp. 370–376, 2015.

[21] W. Nawrot and K. Malecha, "Additive manufacturing revolution in ceramic microsystems," Microelectronics International, vol. 37, no. 2, pp. 79–85, 2020.

[22] M. Dehurtevent, L. Robberecht, J.-C. Hornez, A. Thuault, E. Deveaux, and P. Béhin, "Stereolithography: A new method for processing dental ceramics by additive computer-aided manufacturing," Dental materials, vol. 33, no. 5, pp. 477–485, 2017.

[23] T. Chartier, A. Badev, Y. Aboulaitim, P. Lebaudy, and L. Lecamp, "Stereolithography process: influence of the rheology of silica suspensions and of the medium on polymerization kinetics–cured depth and width," J Eur Ceram Soc, vol. 32, no. 8, pp. 1625–1634, 2012.

[24] Q. Lian, W. Sui, X. Wu, F. Yang, and S. Yang, "Additive manufacturing of ZrO2 ceramic dental bridges by stereolithography," Rapid Prototyp J, vol. 24, no. 1, pp. 114–119, 2018.

[25] H. B. Musgrove, M. A. Catterton, and R. R. Pompano, "Applied tutorial for the design and fabrication of biomicrofluidic devices by resin 3D printing," Anal Chim Acta, vol. 1209, p. 339842, 2022.

[26] W. Du, X. Ren, Z. Pei, and C. Ma, "Ceramic binder jetting additive manufacturing: a literature review on density," J Manuf Sci Eng, vol. 142, no. 4, p. 040801, 2020.

[27] W. Du, J. Roa, J. Hong, Y. Liu, Z. Pei, and C. Ma, "Binder jetting additive manufacturing: Effect of particle size distribution on density," J Manuf Sci Eng, vol. 143, no. 9, p. 091002, 2021.

[28] X. Shen, M. Chu, F. Hariri, G. Vedula, and H. E. Naguib, "Binder jetting fabrication of highly flexible and electrically conductive graphene/PVOH composites," Addit Manuf, vol. 36, p. 101565, 2020.

[29] P. F. Manicone, P. R. Iommetti, and L. Raffaelli, "An overview of zirconia ceramics: basic properties and clinical applications," J Dent, vol. 35, no. 11, pp. 819–826, 2007.

[30] B. Feng, Y. Xin, Z. Sun, H. Yu, J. Wang, and Q. Liu, "On the rule of mixtures for bimetal composites," Materials Science and Engineering: A, vol. 704, pp. 173–180, 2017.